# Acceleration in a Periodic Magnetic Field

M.A. Kutlan

Institute for Particle & Nuclear Physics, Budapest, Hungary

kutlanma@gmail.com

It is shown that acceleration of particles in a homogeneous magnetic field that varies periodically with time (Alfven magnetic pumping) reduces to a diffusion of the particles in momentum space; a connection is established between the diffusion coefficient and the turbulence spectrum $dh^2/dk$ (k-wave number).

## 1. INTRODUCTION

If a slow periodic time variation of a uniform magnetic field B in a turbulent plasma is accompanied by conservation of the adiabatic invariants $p_z = const$, $p_\perp^2/B = const$ ($p_\perp$ and $p_z$ are the particle-momentum components perpendicular and parallel to the field), then the total momentum p of the charged particles can increase exponentially relative to the time t as a result of betatron acceleration and non-adiabatic scattering by the hydromagnetic turbulence (the Alfven magnetic pumping) [l-8]. In a denser plasma, the role of the scattering can be assumed by particle collisions [6,7].

At the present time, in connection with the discovery of pulsars and of the alternating magnetic fields corresponding to them, interest has been renewed in magnetic pumping as one of the possible mechanisms for accelerating charged particles in the atmosphere of a pulsar [9].

As applied to a single particle, Alfven obtained the following equation for the rate of growth of its energy in time:

$$dp/p = dt/\tau, \quad \tau = T/\ln\delta, \quad \delta = p_1/p_0,$$

where $p_0$ and $p_1$ are the values of the total momentum before and after one cycle T of the variation of B [2,5,6], Schluter [7] developed the Alfven analysis further, proving that the character of the periodic variations of the field is of no principal significance. He has shown
in this paper that in the particular case of harmonic variation of the field

$$B = B_0(1 + \beta \cos Wt), \quad \beta < 1 \qquad (1)$$

the effect of acceleration for an ensemble of particles having the same energy is maximal at $\chi =$ W, where $\chi$ is the particle collision frequency.

However, in none of the cited papers was a kinetic equation introduced for the particle distribution function f(p, t) under magnetic-pumping conditions. Considerations to the effect that Alfven acceleration can explain the observed spectrum of the cosmic rays $p^{-2,6}$ are advanced in [4,5], but these considerations are not based on the kinetic equation and are more readily qualitative in character.

In none of the cited papers is there a discussion of the source of the turbulent pulsations that ensure the turbulence intensity needed for the acceleration.

We show in the present paper that the kinetic equation for the distribution function $\bar{f}(p,t)$ averaged over the fast turbulent pulsations can be reduced in the quasilinear approximation, in the presence of a force $F = 1/2(p_\perp/B)dB/dt$, to the diffusion equation in momentum space, with a diffusion coefficient D(p) that depends on the parameters of the alternating magnetic field and on the turbulence spectrum $\Phi(k) = dh^2/dk$ ($h^2$ is the turbulence intensity and k is the wave number (see Sec. 2)). As is well known [10], in the diffusion approximation the turbulence spectrum determines the rate of growth (or damping) of the plasma oscillations, which in turn depends on $\bar{f}(p,t)$, so that the problem of finding the spectrum of the waves and of the particles should be formulated in a self-consistent manner.

In the present paper (see Sec. 3) we propose to obtain a self-consistent system of equations for $\bar{f}(p,t)$ and $\Phi(k)$ by using the equation of the boundary of the cyclotron instability $\gamma = 0$ ($\gamma$ is the increment), which arises in an alternating field as a result of the anisotropy of the angular distribution with respect to the velocities. As is well known, an instability of this type is due to resonance between the waves and the particles at the Larmor frequency, with account taken of the Doppler effect [11]. These results differ from those in spatial periodic magnetic field [14-79].

## 2. DERIVATION OF ACCELERATION EQUATION

In the presence of an alternating magnetic field B(t) that varies with conservation of the adiabatic invariants, $p_z = const$ and $p_\perp^2/B(t) = const$, the equation for the distribution function of the fast particles f(p, t), averaged over the turbulence pulsations, can be written in the quasilinear approximation, as follows from [13,15], in the form

$$\frac{\partial f}{\partial t} + p\sin\theta \frac{\dot{B}}{2B}\left(\frac{\partial f}{\partial p}\sin\theta + \frac{\partial f}{\partial \theta}\frac{\cos\theta}{p}\right) = \widehat{S}f, \qquad (2)$$

where

$$\widehat{S} = \frac{e^2}{mc^2}\frac{1}{\varepsilon\sin\theta}\frac{\partial}{\partial\theta}\sin\theta[P(\theta)]\frac{\partial}{\partial\theta} \qquad (3)$$

is the scattering operator

$$P(\theta) = \frac{\Phi(\omega_B/cp\cos\theta)}{cp\cos\theta},$$

p and $\varepsilon$ : are the dimensionless momentum and energy of the particle, expressed in units of me and $mc^2$, respectively, $\omega_B = eB/mc$ ( e is the charge and m the rest mass of the particle), and $\Phi$ is the spectral function of the turbulence. In writing down (2) we took into account the facts that $p_\perp = p\sin\theta$ and $p_z = p\cos\theta$ ($\theta$ is the angle between p and B), and we assumed that the main turbulence scale L is much larger than the Larmor radii of the fast particles.

The right-hand side in (2) corresponds to the one-dimensional model of cyclotron resonance (to waves traveling along the field, generally speaking, in two directions), and is the zeroth approximation in the expansion of the quasilinear term in powers of $c_A^2/v^2$ ($c_A = B_0/\sqrt{4\pi\rho}$ cA = is the Alfven velocity, $\rho$ is the density of the cold plasma, and $v = cp/\varepsilon$ is the particle velocity). In this approximation, as shown in [13], the resonant interaction of the waves and particles does not lead to turbulent acceleration, and reduces to pure scattering.

We shall solve Eq. (2) in a quasilinear approximation, averaging over the period $T = 2\pi/W$ of the variation of the magnetic field (1). To this end we represent the solution in the form

$$f(\rho,\theta,t) = \bar{f}(p,t) + f_1(\rho,\theta,t) + f_2(\rho,\theta,t) \qquad (4)$$

where $f_1$ and $f_2$ are increments of $\bar{f}$, symmetrical and asymmetrical with respect to $\theta = \pi/2$ (the superior bar denotes averaging over the period T). By virtue of the general properties of scattering that leads to an isotropic distribution [12,13], the average of the symmetrical function over the solid angle, $\langle f_1 \rangle$, is equal to zero, i.e.,

$$\langle f_1 \rangle = \frac{1}{4\pi}\int_0^{2\pi}d\varphi\int_0^{\pi}\sin\theta f_1 d\theta = 0. \qquad \{5\}$$

The reason for breaking down the angle part into $f_1$ and $f_2$ is that according to [13] the isotropization for an initial symmetrical distribution is different than that for an asymmetrical one, so that

$$\hat{S} f_1 \neq 0, \quad \hat{S} f_2 = 0. \tag{6}$$

Using expression {3} for the scattering operator, we can readily show that

$$\langle \hat{S} f_1 \rangle = \frac{e^2}{2\varepsilon m^2 c^2} \left\{ \sin\theta |P(\theta)| \frac{\partial f_1}{\partial \theta} \right\} \Big|_0^\pi = 0. \tag{7}$$

Indeed, the turbulence intensity $h^2 \sim k \Phi(k)$ is limited at $\theta = 0$ and $\pi$, since, by assumption, the main scale of the turbulence is much larger than the Larmor radii of the fast particles, so that $h^2$ decreases with increasing $k = \omega_B / cp\cos\theta$, and the value of $\partial f_1 / \partial \theta$ along the magnetic field is also bounded or else approaches infinity no faster than $1/\sin$ [13].

The linearized kinetic equation for $f_1 + f_2$, which can readily be obtained with the aid of (2) and (4), turns out, after averaging over the solid angle and after using (5) - (7), to depend only on $f_2$:

$$\left\langle \frac{\partial f_2}{\partial t} \right\rangle + p \langle \sin^2\theta \rangle \frac{\partial \bar{f}}{\partial p} \frac{\dot{B}}{2B} = 0,. \tag{8}$$

In the analysis that follows, we confine ourselves to the case of strong scattering, when the isotropization time $\tau_0$ (the time of scattering of a particle with momentum p through an angle 1T) is much smaller than the period T. Subtracting (8) from the unaveraged (over the solid angle) linearized kinetic equation and assuming that in the case of strong scattering we have

$$\frac{\partial f_1}{\partial t} + \frac{\partial f_2}{\partial t} - \left\langle \frac{\partial f_2}{\partial t} \right\rangle \approx 0,$$

and that this equality can be violated only within negligibly short time intervals $\tau_0 \ll T$, we obtain the following equation for $f_1$:

$$p \langle \sin^2\theta \rangle \frac{\partial \bar{f}}{\partial p} \frac{\dot{B}}{2B} \left( \sin^2\theta - \langle \sin^2\theta \rangle \right) = \hat{S} f_1. \tag{9}$$

The strong-scattering approximation is an expansion in powers of the parameter $\alpha / \alpha_0$, where $\alpha$ is the true anisotropy at the plasma stability limit, and $\alpha_0$ is the anisotropy occurring in an alternating field when no account is taken of instability and scattering. The conditions under which this parameter is small will be given later (see Sec. 3) in the analysis of the oscillation increment connected with the anisotropy.

Although the condition $\tau_0 \ll T$ is a definite limitation, at the same time, in accord with [6] (p.80), it is precisely under this condition that the Alfven cycle leads to acceleration, so that the use of this approximation is perfectly justified.

Using (3) and recognizing that ($\langle \sin^2 \theta \rangle = 1/2$, and integrating (9) twice with respect to $\theta$, we obtain

$$f_1 = \frac{\dot{B}}{2B} p \frac{\partial \bar{f}}{\partial p} Q(p,\theta), \qquad (10)$$

where

$$Q(p,\theta) = -\frac{1}{3}\frac{m^2 c^2}{e^2} \varepsilon \int \frac{\cos\theta \sin\theta d\theta}{|P(\theta)|}. \qquad (11)$$

To obtain an equation for $\bar{f}(p,t)$, we substitute (4) in (2), average over the period T, and take into account the fact that $\langle \dot{B} f_2 / 2B \rangle = 0$, since, according to (8), $\dot{B}/2B \sim \partial f_2 / \partial t$ at; retaining terms that are quadratic in the pulsations with the period T, we find in the quasilinear approximation

$$\frac{\partial \bar{f}}{\partial t} + p \sin^2 \theta \left\langle \frac{\dot{B}}{2B} \frac{\partial f_1}{\partial p} \right\rangle + \sin\theta \cos\theta \left\langle \frac{\dot{B}}{2B} \frac{\partial f_1}{\partial \theta} \right\rangle = 0. \qquad (12)$$

Substituting (10) in (12), averaging (12) over the solid angle, and recognizing that according to (6) and (10) we have .

$$\int_0^\pi \sin\theta Q(p,\theta) d\theta = 0, \qquad (13)$$

we obtain for f ( p) after simple transformations an acceleration equation of the diffusion type,

$$\frac{\partial \bar{f}}{\partial t} = \frac{1}{p^2} \frac{\partial}{\partial p}\left[ p^2 D(p) \frac{\partial \bar{f}}{\partial p} \right], \qquad (14)$$

where

$$D(p) = \frac{m^2 c^2 p^2}{6e^2} \left\langle \left(\frac{\dot{B}}{2B}\right)^2 \right\rangle \int_0^\pi \sin^3\theta \left[ \int \frac{\cos\theta \sin\theta d\theta}{|P(\theta)|} \right] d\theta. \tag{15}$$

### 3. FORMULATION OF THE SELF-CONSISTENT PROBLEM OF THE WAVE AND PARTICLE SPECTRUM

Owing to the scattering of the particles by the plasma turbulence, part of the momentum accumulated during the time that the magnetic field (1) is increased as a result of the betatron acceleration ($p_\perp \sim B$), is transferred to the parallel component of the momentum $p_z$. As a result, if the scattering time is small ($\tau_0 \ll T$), the loss of momentum during the time of decrease of the magnetic field (1) turns out to be smaller than the increase of the momentum during the time of the increase of the magnetic field, and the natural result is acceleration (see[6], p. 80 ).

One can expect the cyclotron instability connected with the anisotropy of the angular distribution with respect to the velocities to be, in the presence of an turbulence needed for effective. acceleration. Such an anisotropy in an initially isotropic plasma is the result of the conservation of the adiabatic invariant $p_\perp / B = const$, and is connected with the two-dimensional contraction (expansion) of the Larmor orbits of the particle as a result of the periodic increase (decrease) of the magnetic field. This instability, due to cyclotron resonance between the waves and the particles at the Larmor frequency, occurs at rather low anisotropy when account is taken of the Doppler effect and at sufficiently high particle velocity [11].

According to [11, 10] (seep. 188 of [10]), the expression for the increment of waves having circular polarization and traveling along the field is

$$\gamma = -\frac{2\pi^2 e^2}{c^2 m k^2} \frac{\partial \omega}{\partial k} \int_0^\infty \left\{ f \pm \frac{\omega_B}{\omega\varepsilon} \left( \frac{\partial f}{\partial p_\perp^2} - \frac{\partial f}{\partial p_z^2} \right) p_\perp^2 \right\} dp_\perp^2,$$
$$|p_z| \approx \frac{\omega_B}{ck}, \tag{16}$$

where the upper sign corresponds to the Alfven wave and the lower to the fast magnetosonic wave. The quantity *a,* which characterizes the anisotropy of the distribution function and which determines the properties of the increment (16), is given by

$$\alpha = \frac{\partial f}{\partial p_\perp^2} - \frac{\partial f}{\partial p_z^2} = \frac{1}{2p^2 \sin\theta \cos\theta} \frac{\partial f}{\partial \theta}. \tag{17}$$

Equation (17) can readily be proved if it is recognized that $p_\perp = p\sin\theta \quad p_z = p\cos\theta$.

Were there no scattering (no reaction of the waves on the particles), the corresponding anisotropy $\alpha_0$ would be due only to the adiabatic variation of the momentum in the alternating field (a), and would be very large compared with the anisotropy $a$ in the presence of strong scattering. Thus, in the case of strong scattering, when the angular distribution function is close to isotropic, there exists a small parameter $\alpha/\alpha_0 \ll 1$ in terms of which one can carry out the expansion, as already mentioned in the preceding section.

In the absence of scattering and at sufficiently high concentration of the fast particles, the increment (16) can periodically assume large positive values $\gamma_0 \gg W$. When $\gamma_0 \gg W$, the time $t_0 \sim 1/\gamma_0$ needed for the oscillations to increase by a factor e is much shorter than the period T = 2rr/W, and it is precisely under this condition that the scattering can occur within a time $\tau_0 \ll T$. In other words, the condition $\gamma_0 \gg W$ is necessary for acceleration to be possible, for otherwise, when the instability develops slowly, the scattering of the particles does not lead to complete isotropization at the end of the cycle, and the Alfven acceleration may stop.

The scattering of the particles by the plasma turbulence leads to isotropization of the particle velocities and consequently to a damping of the resonant oscillations, until the magnetic pumping leads to a new cycle of energy redistribution. It can be assumed that the resultant oscillation increment $\gamma(\omega)$, describing the dynamics of the joint action of the magnetic pumping and of the resonance, is equal to zero under strong scattering conditions, i.e., $\gamma(\omega) = 0$. In fact, within short time intervals ($\tau_0 \ll T$) the increment depends on the time and differs from zero. We shall neglect this dependence, however, since we assume the strong scattering condition to be satisfied.

The equation for the plasma instability limit $\gamma(\omega) = 0$ will be used below to obtain self-consistent equations for the wave and particle spectrum. From (5) - (12) it follows that the angle function $f_2$ is not connected with the resonant interaction of the waves and particles, and makes no contribution to the dynamics of the acceleration in the case of strong scattering; we shall therefore take into account in the expression for the increment only that part of the anisotropy (17) which is connected with the angle function $f_2$. Taking this circumstance into account, using expressions (17), (10), and (11), and changing over in (16) to integration with respect to p, we find that the equation $\gamma = 0$ takes the form

$$\int_{\alpha_B/ck}^{\infty} \left\{ f \mp \frac{\omega_B}{\omega \varepsilon} \frac{m^2 c^2}{6e^2} \frac{\dot{B}}{2B} |P(\theta)|^{-1} \sin^2 \theta \right\} p\,dp = 0,$$
$$\cos\theta = \omega_B / ckp.$$
(18)

Eq. (18) can easily be solved with respect to $\Phi(k)$, if it is recognized that at resonance ($\cos\theta = \omega_B / ckp$) the function

$$\Phi(\omega_B / ckp \cos\theta) \equiv \Phi(k)$$

is independent of p. Taking $\Phi(k)$ outside the integral sign in (18), we obtain after simple transformations

$$\Phi(k) = \mp \frac{1}{3} \frac{m^2 c^2 \omega_B}{e^2 \omega k} \frac{\dot{B}}{2B} \left\{ 1 + \lim_{p \to \infty} \left( \frac{\omega_B^2}{c^2 k^2} - p^2 \right) \bar{f}(p) / 2 \int_{\omega_B/ck}^{\infty} p \bar{f} dp \right\}. \qquad (19)$$

Equations (19), (14), and (15) constitute the sought self consistent system of equations for determining the particle spectrum defined by the distribution function f, and the wave spectrum $\Phi(k)$.

**REFERENCES**

1. H. Alfven, Phys. Rev., 75, 1732 (1949).
2. H. Alfven, Phys. Rev., 77, 375 (1950).
3. H. Alfven, Tellus, 6, 232 (1954).
4. H. Alfven, E. Astrom., Nature, 181, 330 {1958).
5. H. Alfven, Tellus, 11, 106 (1959).
6. H. Alfven, K.-H. Felthammer, Cosmic Electrodynamics, 1967, M. F. Bakhareva, V. N. Lomonosov, B. A. Tverskoi, Zh. Eksp. Teor. Fiz. **59**, 2003 (1970).
7. A. Schluter, Zs. Naturforsch, 12a, 822 {1957).
8. V. L. Ginzburg, S. I. Syrovatski'i, Origin of Cosmic Rays, AN SSSR, 1963.
9. V. L. Ginzburg, V. V. Zheleznyakov, V. V. Zaltsev, Sov. Phys.Usp. 12, 378 (1969).
10. 10 A. A. Vedenov, Problems of Plasma Theory, No.3, 203 (1963).
11. R. Z. Sagdeev and V. D. Shafranov, Sov. Phys.-JETP 12, 130 (1961).
12. B. A. Tversko'l, ibid. 25, 317 (1967).
13. B. A. Tversko'i, ibid. 26, 821 (1968).
14. Ashot H. Gevorgyan, K. B. Oganesyan, Edik A. Ayryan, Michal Hnatic, Yuri V. Rostovtsev, Gershon Kurizki, arXive:1703.07637.
15. L.A.Gabrielyan, Y.A.Garibyan, Y.R.Nazaryan, K.B. Oganesyan, M.A.Oganesyan,

M.L.Petrosyan, A.H. Gevorgyan, E.A. Ayryan, Yu.V. Rostovtsev, arXiv:1701.00916 (2017).

16. M.L. Petrosyan, L.A. Gabrielyan, Yu.R. Nazaryan, G.Kh.Tovmasyan, K.B. Oganesyan, A.H. Gevorgyan, E.A. Ayryan, Yu.V. Rostovtsev, arXive:1704.04730.
17. A.S. Gevorkyan , K.B. Oganesyan , E.A. Ayryan , Yu.V. Rostovtsev, arXive:1706.03627.
18. A.S. Gevorkyan, K.B. Oganesyan, E.A. Ayryan, Yu.V. Rostovtsev, arXive:1705.09973.
19. Fedorov M.V., Oganesyan K.B., Prokhorov A.M., Appl. Phys. Lett., **53**, 353 (1988).
20. Oganesyan K.B., Prokhorov A.M., Fedorov M.V., Sov. Phys. JETP, **68,** 1342 (1988).
21. Oganesyan KB, Prokhorov AM, Fedorov MV, Zh. Eksp. Teor. Fiz., **53**, 80 (1988).
22. E.A. Ayryan, K.G. Petrosyan, A.H. Gevorgyan, N.Sh. Izmailian, K.B. Oganesyan, arXive:1703,00813.
23. Edik A. Ayryan, Karen G. Petrosyan, Ashot H. Gevorgyan, Nikolay Sh. Izmailian, K. B. Oganesyan, arXive:1702.03209.
24. D.N. Klochkov, A.H. Gevorgyan, K.B. Oganesyan**,** N.S. Ananikian, N.Sh. Izmailian, Yu. V. Rostovtsev, G. Kurizki, arXiv:1704.06790 (2017).
25. K.B. Oganesyan, J. Contemp. Phys. (Armenian Academy of Sciences), **52,** 91 (2017).
26. AS Gevorkyan, AA Gevorkyan, KB Oganesyan, Physics of Atomic Nuclei, **73**, 320 (2010).
27. D.N. Klochkov, AI Artemiev, KB Oganesyan, YV Rostovtsev, MO Scully, CK Hu, Physica Scripta, **T140,** 014049 (2010).
28. K.B. Oganesyan, M.L. Petrosyan, YerPHI-475(18) – 81, Yerevan, (1981).
29. Petrosyan M.L., Gabrielyan L.A., Nazaryan Yu.R., Tovmasyan G.Kh., Oganesyan K.B., Laser Physics, **17**, 1077 (2007).
30. AH Gevorgyan, KB Oganesyan, EM Harutyunyan, SO Arutyunyan, Optics Communications, **283**, 3707 (2010).
31. E.A. Nersesov, K.B. Oganesyan, M.V. Fedorov, Zhurnal Tekhnicheskoi Fiziki, **56**, 2402 (1986).
32. A.H. Gevorgyan, K.B. Oganesyan, Optics and Spectroscopy, **110**, 952 (2011).
33. A.H. Gevorgyan, M.Z. Harutyunyan, K.B. Oganesyan, E.A. Ayryan, M.S. Rafayelyan, Michal Hnatic, Yuri V. Rostovtsev, G. Kurizki, arXiv:1704.03259 (2017).
34. K.B. Oganesyan, J. Mod. Optics, **61,** 763 (2014).
35. AH Gevorgyan, MZ Harutyunyan, KB Oganesyan, MS Rafayelyan, Optik-International Journal for Light and Electron, Optics, 123, 2076 (2012).

36. D.N. Klochkov, AI Artemiev, KB Oganesyan, YV Rostovtsev, CK Hu, J. of Modern Optics, **57,** 2060 (2010).
37. K.B. Oganesyan, J. Mod. Optics, **62,** 933 (2015).
38. K.B. Oganesyan. Laser Physics Letters, **12**, 116002 (2015).
39. GA Amatuni, AS Gevorkyan, AA Hakobyan, KB Oganesyan, et al, Laser Physics, **18,** 608 (2008).
40. K.B. Oganesyan, J. Mod. Optics, **62,** 933 (2015).
41. A.H. Gevorgyan, K.B.Oganesyan, E.M.Harutyunyan, S.O.Harutyunyan, Modern Phys. Lett. B, **25**, 1511 (2011).
42. A.H. Gevorgyan, M.Z. Harutyunyan, G.K. Matinyan, K.B. Oganesyan, Yu.V. Rostovtsev, G. Kurizki and M.O. Scully, Laser Physics Lett., **13,** 046002 (2016).
43. A.I. Artemyev, M.V. Fedorov, A.S. Gevorkyan, N.Sh. Izmailyan, R.V. Karapetyan, A.A. Akopyan, K.B. Oganesyan, Yu.V. Rostovtsev, M.O. Scully, G. Kuritzki, J. Mod. Optics, **56**, 2148 (2009).
44. A.S. Gevorkyan, K.B. Oganesyan, Y.V. Rostovtsev, G. Kurizki, Laser Physics Lett., **12**, 076002 (2015).
45. K.B. Oganesyan, J. Contemp. Phys. (Armenian Academy of Sciences), **50,** 312 (2015).
46. AS Gevorkyan, AA Gevorkyan, KB Oganesyan, GO Sargsyan, Physica Scripta, **T140,** 014045 (2010).
47. AH Gevorgyan, KB Oganesyan, Journal of Contemporary Physics (Armenian Academy of Sciences) **45,** 209 (2010).
48. K.B. Oganesyan, J. Mod. Optics, **61,** 1398 (2014).
49. AH Gevorgyan, KB Oganesyan, GA Vardanyan, GK Matinyan, Laser Physics, **24,** 115801 (2014)
50. K.B. Oganesyan, J. Contemp. Phys. (Armenian Academy of Sciences), **51,** 307 (2016).
51. AH Gevorgyan, KB Oganesyan, Laser Physics Letters **12** (12), 125805 (2015).
52. ZS Gevorkian, KB Oganesyan, Laser Physics Letters **13**, 116002 (2016).
53. AI Artem'ev, DN Klochkov, K Oganesyan, YV Rostovtsev, MV Fedorov, Laser Physics **17**, 1213 (2007).
54. Zaretsky, D.F., Nersesov, E.A., Oganesyan, K.B., Fedorov, M.V., Sov. J. Quantum Electronics, **16**, 448 (1986).
55. K.B. Oganesyan, J. Contemp. Phys. (Armenian Academy of Sciences), **50,** 123 (2015).
56. DN Klochkov, AH Gevorgyan, NSh Izmailian, KB Oganesyan, J. Contemp. Phys., **51,** 237 (2016).

57. K.B. Oganesyan, M.L. Petrosyan, M.V. Fedorov, A.I. Artemiev, Y.V. Rostovtsev, M.O. Scully, G. Kurizki, C.-K. Hu, Physica Scripta, **T140**, 014058 (2010).
58. Oganesyan K.B., Prokhorov, A.M., Fedorov, M.V., ZhETF, **94**, 80 (1988).
59. E.M. Sarkisyan, KG Petrosyan, KB Oganesyan, AA Hakobyan, VA Saakyan, Laser Physics, **18,** 621 (2008).
60. A.H. Gevorgyan, K.B. Oganesyan, R.V. Karapetyan, M.S. Rafaelyan, Laser Physics Letters, **10**, 125802 (2013).
61. K.B. Oganesyan, Journal of Contemporary Physics (Armenian Academy of Sciences) **51,** 10 (2016).
62. M.V. Fedorov, E.A. Nersesov, K.B. Oganesyan, Sov. Phys. JTP, **31,** 1437 (1986).
63. K.B. Oganesyan, M.V. Fedorov, *Zhurnal Tekhnicheskoi Fiziki*, **57**, 2105 (1987).
64. V.V. Arutyunyan, N. Sh. Izmailyan, K.B. Oganesyan, K.G. Petrosyan and Cin-Kun Hu, Laser Physics, **17**, 1073 (2007).
65. E.A. Ayryan, M. Hnatic, K.G. Petrosyan, A.H. Gevorgyan, N.Sh. Izmailian, K.B. Oganesyan, arXiv: 1701.07637 (2017).
66. DN Klochkov, KB Oganesyan, EA Ayryan, NS Izmailian, Journal of Modern Optics **63,** 653 (2016).
67. K.B. Oganesyan. Laser Physics Letters, **13**, 056001 (2016).
68. DN Klochkov, KB Oganesyan, YV Rostovtsev, G Kurizki, Laser Physics Letters **11,** 125001 (2014).
69. K.B. Oganesyan, Nucl. Instrum. Methods A **812,** 33 (2016).
70. M.V. Fedorov, K.B. Oganesyan, IEEE J. Quant. Electr, **QE-21**, 1059 (1985).
71. D.F. Zaretsky, E.A. Nersesov, K.B. Oganesyan, M.V. Fedorov, Kvantovaya Elektron. **13** 685 (1986).
72. A.H. Gevorgyan, K.B. Oganesyan, E.A. Ayryan, M. Hnatic, J.Busa, E. Aliyev, A.M. Khvedelidze, Yu.V. Rostovtsev, G. Kurizki, arXiv:1703.03715 (2017).
73. A.H. Gevorgyan, K.B. Oganesyan, E.A. Ayryan, Michal Hnatic, Yuri V. Rostovtsev, arXiv:1704.01499 (2017).
74. E.A. Ayryan, A.H. Gevorgyan, K.B.Oganesyan, arXive:1611.04094.
75. A.S. Gevorkyan, K.B. Oganesyan, E.A. Ayryan, Yu.V. Rostovtsev, arXive:1705.09973.
76. A.H. Gevorgyan, K.B. Oganesyan, Laser Physics Letters, **15**, 016004 (2018).
77. K.B. Oganesyan, arXive:1611.08774.
78. E.A. Ayryan, A.H. Gevorgyan, N.Sh. Izmailian, K.B. Oganesyan, arXive:1611.06515.
79. I.V. Dovgan, K.B. Oganesyan, arXive:16.110494